\documentclass{vgtc}                          

\graphicspath{{figures/}{pictures/}{images/}{./}} 

\usepackage{times}                     

\usepackage{tabu}                      
\usepackage{booktabs}                  
\usepackage{lipsum}                    
\usepackage{mwe}                       

\usepackage{mathptmx}                  

\onlineid{1346}

\vgtccategory{Research}

\vgtcinsertpkg

\title{Contextualized Dynamic Explanations: A Vision}

\author{Zhicheng Liu\thanks{e-mail: leozcliu@umd.edu}\\ %
        \scriptsize University of Maryland, College Park %
\and Jason H Li \thanks{e-mail: jason@trustedst.com}\\ %
     \scriptsize Trusted Science and Technology, Inc. %
\and Greg Briskin \thanks{e-mail: greg@trustedst.com}\\ %
     \scriptsize Trusted Science and Technology, Inc.}

\abstract{
Asynchronous data-driven explanations often fail because the content and presentation are not tailored to the target audience, and they provide limited opportunities for active audience engagement. We present a vision for Contextualized Dynamic Explanations (\name), an agentic approach to dynamically generating contextualized multi-modal information interfaces for effective data-driven explanations based on an evolving audience model and a predefined communication intent. 
The premise underlying \name is that it is impossible for communicators to anticipate the full range of interactive scenarios involving the target audience. This observation motivates a set of research challenges focused on developing autonomous agents capable of evaluating communication progress, making context-sensitive decisions, and producing effective information representations.
} 


\newcommand{\name}{\textsc{CoDEx}\xspace}

\newcommand{\bpstart}[1]{\vspace{1mm} \noindent{\textbf{#1.}}}

\newcommand{\RQ}[0]{\vspace{1mm} \noindent{\textit{Research Questions.}}}

\begin{document}



\maketitle

\section{Introduction}
Data-driven explanations are often delivered asynchronously, mediated by an artifact. For example, a designer produces an infographic explaining the timeline and events involved in a phishing attack, and employees at a company view it during their security training sessions; a journalist creates an interactive visualization explaining the impacts of a new federal bill, and readers engage with the visualization days later; a sports analyst produces a YouTube video explaining how a basketball team lost a playoff series, and viewers watch at different times. In many such situations, the explanation may fail to be communicated effectively to the audience due to a range of factors. The audience may lack the necessary background knowledge to relate the explanation to their prior understanding. The information design of the artifact may be ineffective, making it difficult for the audience to interpret the data and the intended meaning. Furthermore, the artifact may serve only as a conduit for the communicator’s ideas without supporting audience to explore or request clarification.

To promote more active audience engagement in understanding explanations, a line of work on explorable explanations \cite{bret_victor_explorable_nodate, collaris_explainexplore_2020,dragicevic_increasing_2019} aims to provide interactive interfaces that allow users to access contextual information, model scenarios, and consider alternative possibilities in an explanation. However, explorable explanations do not fully address the sources of communication failure, such as lack of background knowledge and ineffective information design. 

With advances in generative AI, we present our vision for \textit{Contextualized Dynamic Explanations}, an agentic approach to explanation communication. We follow Keil's definition and characterization of explanation \cite{keil_explanation_2006}: 
in contrast to law-like phenomena without mechanisms (e.g., if X, then Y), explanations provide \textit{causal relations or mechanisms} with the aim of making something intelligible to someone who did not understand it before.  A contextual dynamic explanation (\name) goes beyond explorable explanations to incorporate audience diversity and engagement as a fundamental consideration in their design and implementation. A \name is deeply contextualized: the content and presentation of an explanation are personalized for its audience and are adaptive to the dynamics of the interactive process. It is not feasible for the communicator to anticipate in advance all potential questions from, or actions taken by, the audience. Therefore, the content and presentation at each turn have to be dynamically generated. We illustrate \name with two motivating examples (Section 2), and describe the main properties of a \name (Section 3). We then outline the research challenges and opportunities in developing AI agents to enable \name (Section 4). Finally, we discuss how \name relates to and differs from related concepts in the literature (Section 5).
\section{Motivating Scenarios}
The following scenarios illustrate the key capabilities of a \name:

\bpstart{S1} A journalist wants to generate a \name to explain a recent finding that drinking coffee may reduce the risk of type 2 diabetes. The \name should target three types of audience: for the general public, the article can highlight the finding that natural compounds in coffee may improve the way the human body uses insulin and discuss implications of the finding on coffee drinking practice; for healthcare professionals, the article can focus on effect size and clinical implications; for researchers, the article can present information on the study method, technical details, and biochemical mechanisms. 

The audience can interact with the \name in various ways, including — but not limited to — posing follow-up questions (e.g., a general reader asks, ``\textit{What is the function of insulin?}''), requesting detailed information (e.g., a healthcare professional retrieves study data to evaluate the significance of the results), and verifying claims (e.g., a researcher conducts a multiverse analysis on the study data). 

The representation of the information should also be tailored to the audience type: the explanation on the role and function of insulin may be in the form of an animated infographic based on an appropriate analogy or metaphor; the study data may be visualized using statistical charts showing uncertainty information, and the multiverse analysis results are presented using small multiples. 

To adaptively guide this process, the \name monitors signals of audience understanding and trust—such as the types of questions asked, the depth of follow-up requests, or the consistency of user actions. Based on this evolving model of the audience, the system selects the next explanatory step, whether to elaborate, simplify, provide evidence, or summarize, ensuring the communication remains aligned with the audience's needs and the journalist's intent.

\bpstart{S2} A defense system deployed in an organization has identified a security vulnerability in the company network: a potential attacker can access confidential data from a legacy VPN account associated with a third-party vendor. A \name on this finding is generated for the employees in the legal and IT departments. The legal department is primarily interested in whether this identified vulnerability means a compliance violation and whether any breach has already occurred. The IT department is more interested in the potential attack paths and the effects of possible mitigation. 

The interaction between the audience and the \name is dynamic. For instance, the legal team can ask about the causal mechanisms in the explanation (e.g., VPN point $\rightarrow$ file server $\rightarrow$ credential server) and get contextual information (e.g., whether the mere risk of leaving data publicly accessible without any evidence of actual access constitutes a violation under the EU's data protection law GDPR); the IT staff can ask for scenario simulations and perform what-if exploration (e.g., evaluate the effect of blocking SSH from the file server). 

The representation of information and the associated interaction in the \name are also contextualized to the type of content and audience. For instance, there is support to directly manipulate network visualizations showing complex relationships between nodes, applications, and ports to explore what-if scenarios; text visualization techniques can be applied to highlight the similarity and differences between the relevant laws in different jurisdictions to understand the impact in terms of compliance. 

The \name not only reacts to queries, but also proactively guides the interaction. For example, when legal staff focus on compliance concerns, it can suggest examining relevant GDPR precedents or provide summaries of past enforcement cases. Similarly, if IT staff explore one attack path, the \name can recommend simulating additional unconsidered paths or highlight overlooked vulnerabilities.

\section{Properties of a \name}
\begin{figure}[th]
 \centering \includegraphics[width=0.8\linewidth]{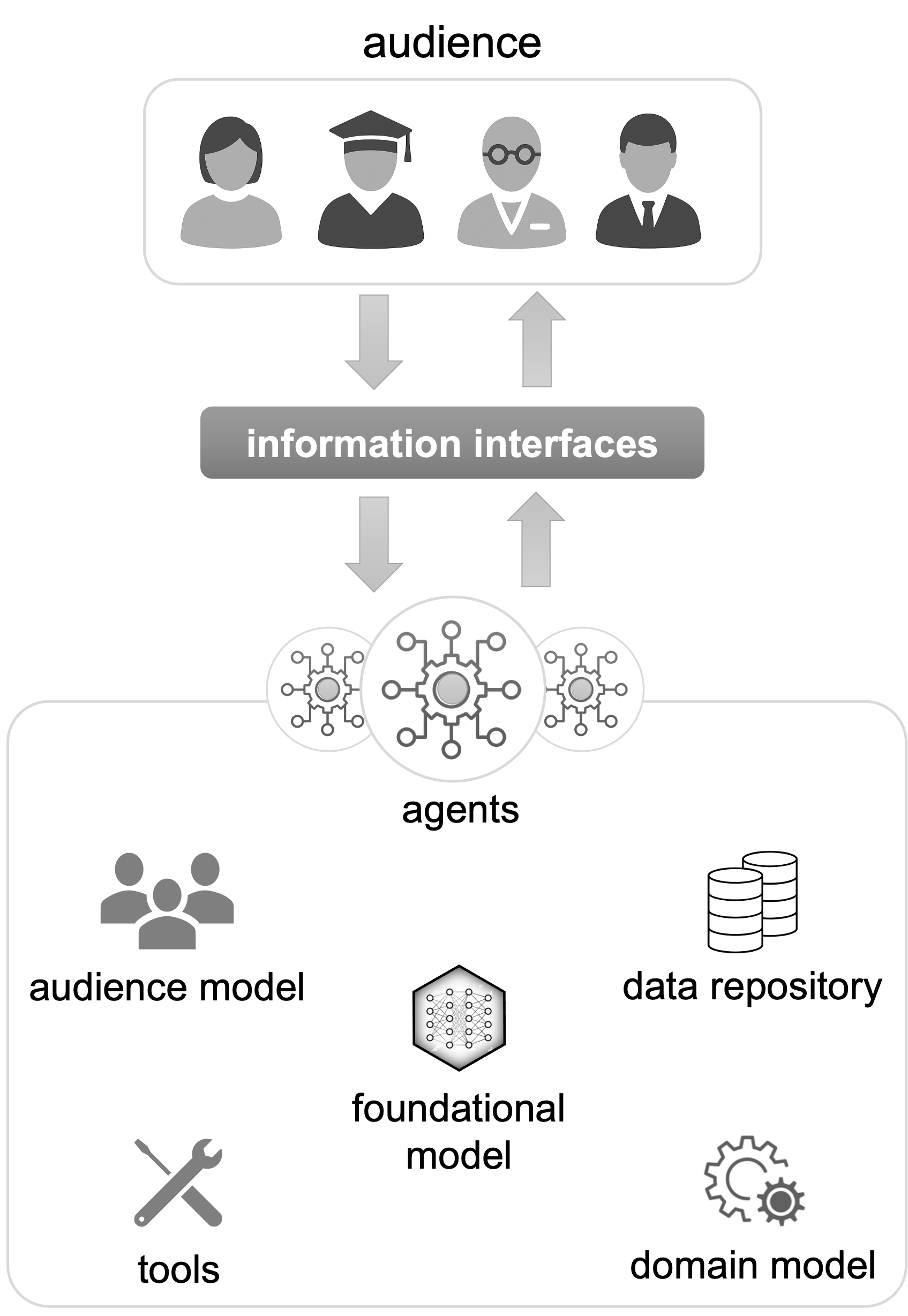}
 \caption{An illustration of the components of a \name \label{fig:codex}}
\end{figure}

A \name has the following essential properties, in addition to the premise that it happens asynchronously between the communicator and the audience, mediated by an artifact, as mentioned in the introduction. 

\bpstart{Communication intent} A \name assumes a clear communication intent, which consists of a well-defined explanation and target audience. An explanation often is crafted by a human communicator (S1); it can also be generated by a computational system (S2). It conveys why events occur or why a statement holds true/false by linking causes and effects, illustrating how components relate, or describing how processes unfold. 

The target audience may be described in terms of their demographics (e.g., occupation, education level), interests, values, and goals, prior knowledge and experience with the topic, behavioral and media habits (e.g., uses mobile much more than PCs), communication preferences (design and style of representations), and accessibility considerations (e.g., neural divergent, visually impaired). 

This requirement of a clearly defined communication intent sets a \name apart from tools designed with the aim of supporting more open-ended activities such as sense making or exploratory data analysis. 

\bpstart{Contextualized information interfaces} The interaction between the audience and a \name is mediated by information interfaces. The information interfaces both present data related to the explanation, and enable the audience to actively engage with the explanation.  They are tailored to the target audience's prior knowledge, interests, experiences, personal traits, and preferences. Such contextualization may be based on audience profiles defined \textit{a priori} or explicit instructions from the audience, and implicitly inferred from the interaction dynamics. In any case, the \name needs to maintain and update a model that describes the target audience. 

\bpstart{Dynamic generation and update} The deep contextualization in a \name implies that \textit{it is impossible to anticipate all the questions and requirements from the audience}. The information interfaces thus have to be dynamically generated and updated, instead of pre-wired. 
For example, in S2, the IT staff may initially require semantic zooming capabilities in a visualization of multiple potential attack paths, and later want to expand or collapse nodes within the same visualization.

\bpstart{Heterogeneous models} An effective \name needs to leverage and coordinate multiple models for dynamic generation and updating of content and presentation. Each model can serve as a knowledge resource and possess descriptive or prescriptive powers. Examples of such models include but are not limited to:

\begin{itemize}
    \item A \textit{foundational model} (e.g., LLM, VLM) trained on internet-scale public knowledge covering diverse topics. It can clarify key terms or concepts related to the explanation to be communicated, provide analogies to make a difficult concept easier to understand, and suggest alternative scenarios among many capabilities.
    
    \item A \textit{domain-specific model} developed to solve problems within a specific domain.  Such a model can be more desirable than a foundational model when it runs on smaller proprietary data, can be easily implemented based on effective rules and procedures, or provides deterministic and explainable outputs. In S2, the defense system used to identify security vulnerabilities has access to private knowledge such as company network configurations, and may function based on formal representations of Cyber Threat Intelligence (CTI) knowledge \cite{noauthor_what_nodate} instead of neural network representations. 
    
    \item An \textit{audience model} describing the traits, knowledge, and preferences of the target audience. The model may be defined using approaches such as segment descriptions (categorizing a person based on attributes such as demographics and job role) and personas (archetypes that represent key audience segments). The behavior, knowledge, and preferences of the target audience can also be captured by user models learned from unstructured observations \cite{shaikh2025creating}. 
\end{itemize}

\bpstart{Multi-Modality} The information in a \name is usually presented in multiple modalities. Natural language, whether expressed in writing or speech, is usually the most intuitive medium. Visual representations of concepts, processes, and datasets are often needed as well. The audience's requests can be expressed not only using natural language, but also through direct manipulation or gesture on visual or tangible interfaces. Furthermore, the use of animation in a video form can also enhance the delivery of the explanation. To generate information interfaces in these modalities, the \name needs to employ various \textit{tools}, such as visualization libraries and UI toolkits.

\section{Research Challenges}
Since it is impossible to anticipate all possible interaction scenarios between a \name and the target audience, the \name must behave as an autonomous agent that evaluates the state of communication, makes decisions on appropriate actions, and presents information in effective representations. 

Although solutions for various sub-problems in \name are emerging and maturing, the properties of \name present a set of new problems that cannot be solved by simply piecing together these existing solutions. For example, LLMs like ChatGPT have introduced more persistent memory \cite{chatgpt_memory} to remember information across conversational sessions. However, such memory features are not sufficient to serve as audience models: they may not be able to distinguish between what users know and what they misunderstand; their models of users' frames of reference (beliefs, prior knowledge, values, goals) are still shallow. Although LLMs can generate code for charts and interactive interfaces, it still often takes multiple rounds of iterations to correctly generate a design, requiring users to interpret and even debug the code. In general, it remains a major challenge to achieve multi-turn continuity in maintaining complex visual states and coherently tracking the status of the explanation process. To build an effective \name, the following research challenges must be addressed.

\subsection{Audience Model} 
A \name needs to have an understanding of the initial state of an explanatory effort. Specifically, who the audience is and what they already know. This understanding should be captured in an audience model, which can be revised and updated throughout the interaction process. To create such models, we may expect verbal description of the type and/or personas of the target audience as input; in some cases, the input may be in the form of a sophisticated audience model learned from empirical data \cite{brusilovsky_user_2007,shaikh2025creating}.

\RQ~How to ingest descriptions of the audience at varying levels of detail in different formats into a \name to build an audience model? What kind of internal representations should be used to capture the audience's existing cognitive frames and knowledge? How can the audience model be integrated into reasoning processes and dynamically updated throughout the interaction?

\subsection{Goal Specification} 
Similar to any AI agent, a \name requires a goal: the successful fulfillment of a communication intent. We consider the following components in a clear goal definition: (1) the explanation to be delivered,  (2) rhetorical strategies, (3) methods to evaluate the outcome of communication, and (4) the target audience. Having discussed the target audience above, we now explore the remaining three components.   

\vspace{2mm}

\bpstart{Explanation} An explanation specification may include the following elements: subject matter (what is being explained), claim (what the explanation asserts), reasoning/mechanism (how or why the claim holds), context/background (prior knowledge or assumptions necessary for understanding), examples/analogies (for illustration), and alternative interpretations/scenarios. Certain elements (e.g., subject matter and claim) are best specified directly by the communicator, whereas others (e.g., context/background) may be generated or refined by \name. 

\RQ~Which of the explanation elements require explicit specification by the communicator, and what aspects can be delegated to agentic generation? What are the implications of generated elements on the explanation outcomes? 

\vspace{2mm}
\bpstart{Rhetorical Strategies} The same set of explanation elements can be structured using different rhetorical strategies. For instance, one can present the claim first, and then justify it with reasoning and evidence. Alternatively, one may build up the case step by step, starting with evidence, and using reasoning to reach the claim. The choice of a strategy determines the initial setup and the interaction flow. Such choices can be made by the communicator, or planned by the agent. 

\RQ~At what granularity should the communicator specify the rhetorical sequencing? How can agents flexibly restructure explanations (e.g., drilling into mechanisms vs. giving high-level overviews) based on the audience model and partial communicator input? What representational frameworks can capture rhetorical strategies at a fine-enough granularity for adaptive explanations? How to balance coherence and adaptivity in mixed-initiative approaches to choosing rhetorical strategies? 

\vspace{2mm}
\bpstart{Assessment Methods} A goal specification would not be complete without methods to assess whether the goal has been achieved. Previous work has suggested that communication intents can be formulated as learning objectives, expressed as pairs of a verb and a noun in a cognitive taxonomy (e.g., recall details, compare theories) \cite{adar_communicative_2020}. This approach is promising, and the learning objectives can be formulated by the communicator or an agent. 

\RQ~How can the formulation of learning objectives be contextualized for different audience profiles? What unobtrusive signals can reliably indicate whether the learning objectives are fulfilled? How should a \name represent and respond to partial understanding or misinterpretation, and when is ``good enough'' sufficient? Besides learning objectives, what other measures (e.g., trust) may indicate successful communication? 

\subsection{Response Generation} 
When the audience asks a question or performs an action through the interface, the \name needs to generate an appropriate response. Specifically, the following decisions need to be made:

\vspace{2mm}
\bpstart{Determine Response Type}
Depending on the audience's request, the response can provide supporting evidence (e.g., show detailed attack paths in S2), retrieve contextual or background information (e.g., GDPR article 33), make an analogy to make easier to understand, or generate a simulation for the audience to explore. 

\RQ~What are the different types of responses in a \name?  How do we determine the appropriate response type based on the request, interaction history, and audience model? 

\vspace{2mm}
\bpstart{Retrieve Relevant Information} A \name has access to multiple knowledge sources, including various models and data repositories. Depending on the audience's question, some sources can provide more relevant and reliable information than others. General questions (e.g., ``\textit{What is the function of insulin?}'') are often best addressed by Large Language Models (LLMs). In contrast, responses that require proprietary data are better served by local repositories and domain-specific models. The \name also needs to take into consideration the state of the current information interfaces in retrieving information for response generation. 

\RQ~How can we determine which knowledge sources to access for the retrieval of relevant information for response generation? If multiple sources are required, how to synthesize and reconcile the information? How to address the hallucination problem in the context of retrieving relevant information?

\vspace{2mm}
\bpstart{Design Effective Information Interfaces} With the retrieved information, the agent must present it using the modalities and representations that best support the communication goal. For instance, when conveying structured data with multiple attributes, it should select visual encodings that most effectively advance understanding. When enabling real-time what-if explorations, it should dynamically generate interactive widgets within the current interface to allow users to adjust parameters on the fly.

\RQ~How can an agent dynamically choose the appropriate modality (e.g., an interactive interface vs. a verbal element) given the context of ongoing interaction? How do current approaches to generating interactive visualizations and user interfaces perform in the context of \name, where a clear communication goal must be achieved through multimodal conversational interfaces?  How to automatically generate effective information interfaces for communication? 

\subsection{Progressing Monitoring and Steering} 
During the interactive process, the \name must evaluate the current state of communication in real time and determine whether the goal has been achieved. This requires access to the audience’s understanding of the explanation. Such understanding can be inferred implicitly from the audience’s actions and questions; however, the information available at runtime may be insufficient to support reliable inferences. Alternatively, the \name can explicitly solicit the audience’s responses through targeted questions or assessments aligned with predefined learning objectives, though this approach risks disrupting the flow of communication. In addition, the audience may convey useful information about their prior knowledge, interests, or personal traits during the process. Such information needs to be incorporated into the audience model. 

Based on the evaluation of the explanation progress, the \name should proactively 
steer the course of explanation to prevent drifts away from the predefined goal. For example, if the audience asks questions that are tangential to the communication intent, or wants to forage additional data on a related but different topic, such activities should be deemed out of the scope of the current explanation. If the \name infers that the audience has already grasped the explanation, it should proceed to conclude the session.

\RQ~What memory structures and representations are needed to maintain explanatory coherence over time? How to develop methods for non-intrusive evaluation of progress and revision of audience model? How to combine explicit and implicit approaches to assess understanding without degrading the user experience? How can an agent decide when to take the lead versus when to follow the audience's cues, and how can it dynamically adapt this balance?

\section{Related Approaches}
\name extends the idea of explorable explanations \cite{bret_victor_explorable_nodate} to incorporate contextualization in light of the advances in generative and agentic AI. The contextualization features draw ideas from personalized visualizations and explanations \cite{adar_persalog_2017,geng_improving_2022,schneider_personalized_2019} and adaptive visualizations \cite{shin_drillboards_2025,hutchison_towards_2012}. A key distinction between \name and the previous work lies in the substantially greater freedom it affords the audience to participate in the explanatory process, both through natural language questions and through direct interaction with the interface. The space of possible interactions in a \name is not predefined. To respond effectively and guide the audience, agentic AI must serve as the backbone of \name. 

Narrative visualizations \cite{segel_narrative_2010}, interactive articles \cite{hohman2020communicating}, data videos \cite{amini_understanding_2015} are different ways to communicate data-driven insights. Some of these techniques are intended for storytelling purposes. While storytelling and explanation are related, they are not the same: explanations prioritize clarity and understanding, whereas stories emphasize engagement and emotional resonance. These techniques can be employed in a \name. However, they generally do not require an agentic framework. 

Explainable AI research \cite{mohseni_multidisciplinary_2021} focuses on designing, developing, and evaluating AI systems whose outputs and behaviors are interpretable. \name is not concerned with developing methods to make AI systems more explainable. If an explainable AI system does not provide a clear answer to explain a phenomenon and requires the user to deduce the answer independently, it falls outside the scope of \name. However, once such an answer is obtained, it can be delivered to the target audience using \name. The main motivation behind \name is to establish a \textit{dialogue} between the audience and the agents responsible for delivering an explanation, enabling the audience to develop an understanding of what leads to an outcome through mixed-initiative communication and what-if exploration.

\section{Conclusion}
Contextualized Dynamic Explanations is a new approach to rethinking how asynchronous explanations can be fundamentally changed by agentic AI. The approach has broad applicability across diverse domains where effective communication of explanations is critical. As a first step to make this approach a reality, we highlight the main properties of a \name, and outline a set of research challenges and opportunities. 

\acknowledgments{
This fundamental research was developed with the support from the Defense Advanced Research Projects Agency (DARPA) under contract HR001125C0056 for the RC-Vis opportunity. The views, opinions and/or findings expressed are those of the authors and should not be interpreted as representing the official views or policies of the Department of Defense or the U.S. Government.}

\bibliographystyle{abbrv-doi}

\bibliography{template}
\end{document}